\documentclass[prb,onecolumn,amsmath,amssymb, nofootinbib, superscriptaddress]{revtex4}
\usepackage[pdftex]{graphicx}% Include figure files
 \usepackage{amsmath}
\usepackage[T1]{fontenc}
\usepackage{flushend}
\usepackage{amssymb}
\usepackage{amsfonts}
\usepackage{bm}
 \usepackage{amsmath} 
\usepackage{lipsum}
\usepackage{amsfonts} 
\usepackage{amssymb, mathrsfs}
\usepackage{braket}
\usepackage{graphicx} 
\usepackage{subfigure}
\usepackage{bbm}
\def\beq{\begin{equation}}
\def\eeq{\end{equation}}
\def\bsp{\begin{split}}
\def\esp{\end{split}}
\def\bea{\begin{eqnarray}}
\def\eea{\end{eqnarray}}
\def\ba{\begin{array}}
\def\ea{\end{array}}

\def\dg{\dagger}

\def\lb{\left(}
\def\rb{\right)}

\def\l.{\left.}
\def\r.{\right.}

\def\ra{\rangle}
\def\la{\langle}

\def\bo{\vec{k}}

\begin{document}

\date{\today}
\title{\large Photoinduced Topological Phase Transitions in  Topological Magnon Insulators }
\email{sowerre@perimeterinstitute.ca}
\author{S. A. Owerre}
\affiliation{Perimeter Institute for Theoretical Physics, 31 Caroline St. N., Waterloo, Ontario N2L 2Y5, Canada.}

\maketitle
%\section{Introduction}
\noindent\textbf {Topological magnon insulators are the bosonic analogs of electronic  topological insulators. They are manifested in  magnetic materials with topologically nontrivial magnon bands as realized experimentally in a quasi-two-dimensional (quasi-2D)  kagom\'e ferromagnet Cu(1-3, bdc), and they also possess protected magnon edge modes. These topological magnetic materials can transport heat as well as spin currents, hence they can be useful  for spintronic applications. Moreover, as magnons are charge-neutral spin-${\bf 1}$ bosonic quasiparticles with a magnetic dipole moment, topological magnon materials can also interact with electromagnetic fields through the Aharonov-Casher effect.    In this report, we study photoinduced topological phase transitions in intrinsic topological magnon insulators in the kagom\'e ferromagnets. Using magnonic Floquet-Bloch theory, we show that by varying the  light intensity, periodically driven intrinsic topological magnetic materials can be manipulated into different topological phases with different sign of the Berry curvatures and the thermal Hall conductivity. We further show that,  under certain conditions, periodically driven gapped topological magnon insulators can also be tuned to synthetic gapless topological magnon semimetals with Dirac-Weyl magnon cones.  We envision that this work will pave the way for  interesting new potential practical applications in topological magnetic materials.}

\vspace{10px}

Topological insulators have captivated the attention of researchers in recent years and they  currently represent one of the active research areas  in condensed matter physics \cite{top1, top2,top3,top4,top5}. These nontrivial insulators can be realized in  electronic systems with strong spin-orbit coupling and a nontrivial gap in the energy band structures. They also  possess Chern number or $\mathbb{Z}_2$ protected metallic edge or surface modes that can transport information without backscattering \cite{top5}.   In principle, however, the ubiquitous notion of topological band theory  is independent of the statistical nature of the quasiparticle excitations. In other words, the concept of Berry curvature and Chern number can be defined for any topological band structure irrespective of the quasiparticle excitations. Consequently, these concepts have been extended to bosonic systems with charge-neutral quasiparticle excitations such as magnons \cite{alex0, zhh,alex4, alex4h,cao,sm,rold,sol,sol1,alex5b,alex5c,ruck, cherny, swang, kova,alex4a, pant}, triplons \cite{rom,mcc}, phonons \cite{pho1,pho3}, and photons \cite{ling}.

Topological magnon insulators \cite{alex0, zhh,alex4, alex4h,cao,sm,rold,sol,sol1,alex5b,alex5c, alex4a} are the bosonic analogs of electronic  topological insulators. They  result from the nontrivial low-energy excitations of insulating quantum magnets with spin-orbit coupling or Dzyaloshinskii-Moriya (DM) interaction \cite{dm,dm2}, and exhibit  topologically nontrivial magnon bands and  Chern number protected magnon edge modes,  with similar properties to those of electrons in topological insulators \cite{top1, top2,top3,top4,top5}. Theoretically, topological magnetic excitations can arise in different lattice geometries with DM interaction, however their experimental observation  is elusive in real magnetic materials. Recently, intrinsic topological magnon insulator has been observed experimentally in a quasi-2D  kagom\'e ferromagnet Cu(1-3, bdc) \cite{alex5b}. Moreover, recent evidence of topological triplon bands have also been reported in a dimerized quantum magnet SrCu$_2$(BO$_3$)$_2$ \cite{rom,mcc}. These magnetic materials have provided an interesting transition from electronic  to bosonic topological insulators.

Essentially, the intrinsic properties of a specific topological magnon insulator are material constants that cannot be tuned, thereby hindering a topological phase transition in the material.  In many cases of physical interest, however, manipulating the intrinsic properties of topological magnetic materials could be a stepping stone to promising practical applications, and could also provide a platform for studying new interesting features such as  photo-magnonics \cite{benj}, magnon spintronics \cite{magn, benja},  and ultrafast optical control of magnetic spin currents \cite{ment, tak4, tak4a,walo}.  One way to achieve these scientific goals is definitely through light-matter interaction induced by photo-irradiation. In recent years, the formalism of photo-irradiation  has been a subject of intensive investigation  in electronic systems such as graphene and others \cite{foot3,foot4,foot5,gru,fot,fot1,jot,fla,we1,we2,we3,we4,we5,we6, gol,buk,eck1,ste, ple,ew, dik, lin, du,du1, delp, eza, zhai, saha,roy,roy1}. Basically, photo-irradiation allows both theorists and experimentalists to engineer topological phases from trivial systems and also induce photocurrents and phase transitions in topologically nontrivial systems. In a similar manner to the notion of bosonic topological band theory, one can also extend the mechanism of photo-irradiation  to bosonic systems.
  
  In this report, we theoretically investigate photo-irradiated intrinsic topological magnon insulators in the kagom\'e ferromagnets and their associated topological phase transitions. One of the main objectives of this report is to induce tunable parameters in intrinsic topological magnon insulators, which subsequently drive the system into a topological phase transition. We achieve this objective by utilizing the quantum theory of magnons, which are charge-neutral spin-$1$ bosonic quasiparticles   and carry  a magnetic dipole moment.  Therefore, magnons can couple to  both time-independent\cite{ mei, mei1,xr,xr1} and periodic time-dependent (see Methods) \cite{ow1} electric fields through the  Aharonov-Casher (AC) effect \cite{aha}, in the same manner that electronic charged particles couple through the Aharonov-Bohm (AB) effect \cite{aha1}. Quite distinctively, for the periodic time-dependent electric fields (see Methods) \cite{ow1}, this results in a periodically driven  magnon system, and thus can be studied by the  Floquet-Bloch theory.   Using this formalism, we show that intrinsic topological magnon insulators can be tuned from one topological magnon insulator to another with different Berry curvatures, Chern numbers, and thermal Hall conductivity. Moreover, we show that, by manipulating the light intensity, periodically driven intrinsic topological magnon insulators can also transit to synthetic gapless topological magnon semimetals.  Therefore,  the magnon spin current in topological magnetic materials can be manipulated by photo-irradiation, which could be  a crucial step towards potential practical applications.
  
\vspace{10px}
\noindent \textbf{\large Results}
  
\begin{figure}[!]
\includegraphics[width=5in]{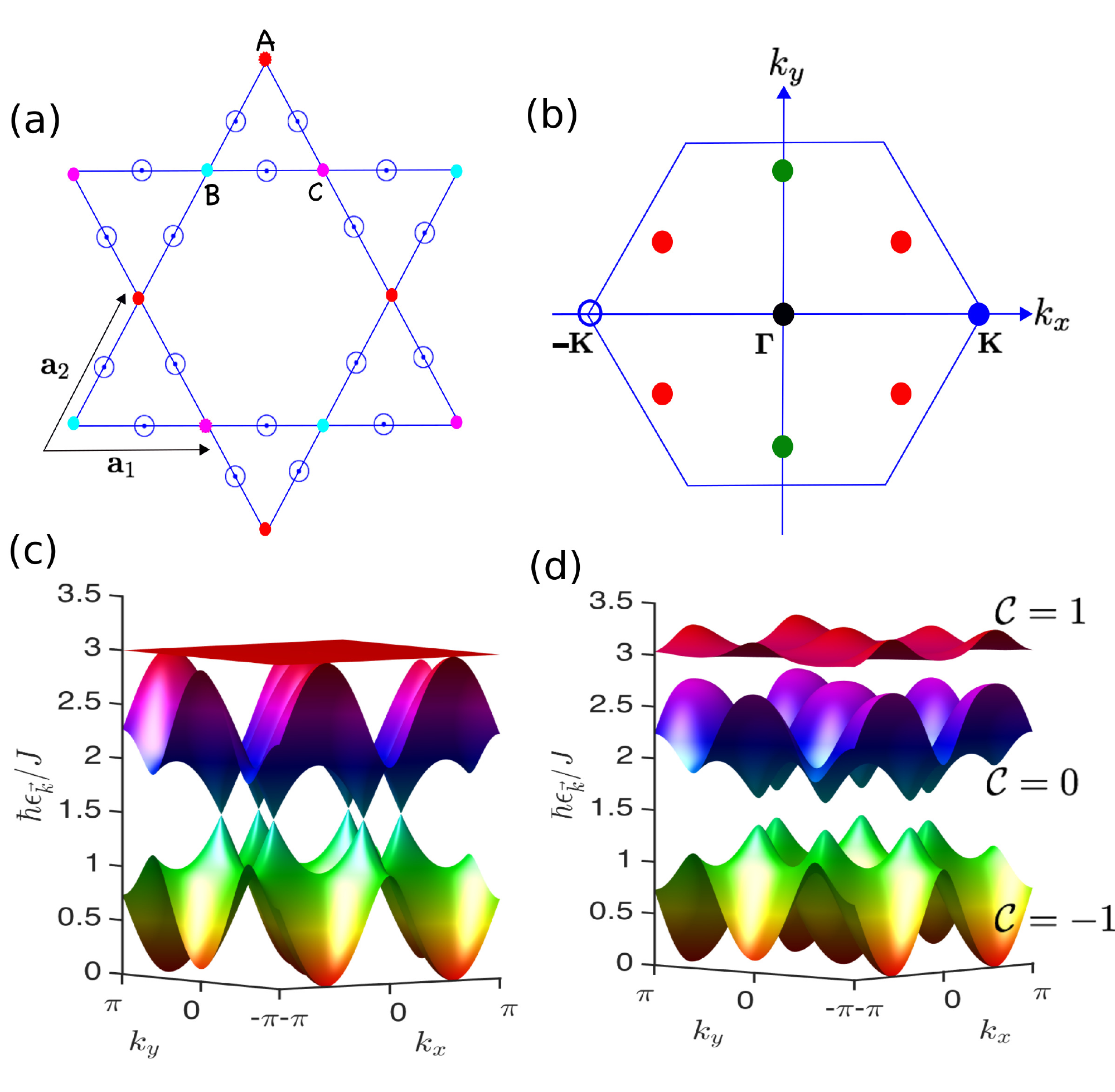}
\caption{(a) Schematic of the kagom\'e lattice with three sublattices A, B, C as indicated  by coloured dots, and the out-of-plane DM interaction is indicated by open circles. (b) The first Brillouin zone of the kagom\'e lattice with two inequvilaent high symmetry points at $\pm{\bf K}$. The red and green dots denote the photoinduced Dirac-Weyl magnon nodes as will be discussed  later.  (c) Magnon bands of undriven insulating kagom\'e ferromagnets with $D/J=0$, showing Dirac magnon nodes at $\pm{\bf K}$, formed by the two lower dispersive bands. (d) Topological magnon bands of undriven  insulating  kagom\'e ferromagnets with  $D/J=0.15$.}
\label{klat}
\end{figure}

\noindent\textbf{Topological magnon insulators.}~~ We consider the simple microscopic spin Hamiltonian  for intrinsic topological magnon insulators in the kagom\'e ferromagnets \cite{alex5b}
\begin{align}
\mathcal H&=\sum_{\la \ell\ell^\prime \ra } \big[ -J{\vec S}_{\ell}\cdot{\vec S}_{\ell^\prime}+\vec{D}_{\ell\ell^\prime}\cdot ({\vec S}_{\ell}\times{\vec S}_{\ell^\prime})\big]-\vec{B}\cdot\sum_{\ell} \vec{S}_{\ell}.
\label{kham}
\end{align} 
The first summation is taken over nearest-neighbour (NN) sites  $\ell$ and $\ell^\prime$  on  the  2D kagom\'e lattice, and $\vec{D}_{\ell\ell^\prime}$ is the DM vector between the NN sites due to  lack of an inversion center as depicted in Fig.~\eqref{klat}a. The last term is the Zeeman coupling to an external magnetic field  $\vec{B}=g\mu_B\vec{H}$,  where $\mu_B$ is the Bohr magneton and $g$ the spin $g$-factor. Topological magnon insulators \cite{alex0, zhh,alex4, alex4h,cao,sm,rold,sol,sol1,alex5b,alex5c} can be captured by transforming the spin Hamiltonian to a bosonic hopping model using the Holstein-Primakoff (HP) spin-boson transformation.  In this formalism, only the DM vector parallel to the magnetic field contributes to the noninteracting bosonic Hamiltonian \cite{alex5b,alex0}, but other components of the DM vector can be crucial when considering magnon-magnon interactions \cite{cherny}. Here, we limit our study to noninteracting magnon system as it captures all the topological aspects of the system \cite{alex5b,alex0}.  We consider then an external magnetic field along the $z$ (out-of-plane) direction, $\vec{B}=B\hat{z}$, and take the DM vector as $\vec{D}_{\ell\ell^\prime}=D\hat z$. The Holstein-Primakoff (HP) spin-boson transformation is given by $S_{\ell}^{ z}= S-a_{\ell}^\dagger a_{\ell},~S_{\ell}^+\approx \sqrt{2S}a_{\ell}=(S_{\ell}^-)^\dg$, where $a_{\ell}^\dagger (a_{\ell})$ are the bosonic creation (annihilation) operators, and  $S^\pm_{\ell}= S^x_{\ell} \pm i S^y_{\ell}$ denote the spin raising and lowering  operators. Applying the transformation to Eq.~\eqref{kham} yields the  bosonic (magnon) hopping  Hamiltonian 
 \begin{align}
\mathcal H&=-t_0\sum_{\la \ell\ell^\prime \ra} \big(a_{\ell}^\dagger a_{\ell^\prime}e^{i\varphi_{\ell\ell^\prime}} + \text{H.c}\big) +t_z\sum_{\ell}n_\ell,
\label{ham1}
\end{align}
where $n_\ell=a_{\ell}^\dagger a_{\ell}$ is the number operator; $t_0=JS\sqrt{1+(D/J)^2}$ and  $t_z=4JS+B$ with $t_z>t_0$. The phase $\varphi_{\ell\ell^\prime}=\pm\varphi=\pm\tan^{-1}(D/J)$ is the fictitious magnetic flux in each unit triangular plaquette of the kagom\'e lattice \cite{alex0}, in analogy to the Haldane model \cite{top1}.  The Fourier  transform of the magnon Hamiltonian is given by $\mathcal H=\sum_{\vec{k}} \psi_{\vec{k}}^\dg \mathcal H({\vec{k}})\psi_{\vec{k}}$, with $\psi_{\vec{k}}=(a_{\vec{k},1},a_{\vec{k},2},a_{\vec{k},3})^{\text{T}}$, where $\mathcal H(\vec{k})=t_z \text{I}_{3\times 3}- \Lambda(\vec{k})$,
\begin{align}
\Lambda(\vec{k}) &=2t_0
\begin{pmatrix}
0& \cos k_2e^{-i\varphi}& \cos k_3e^{i\varphi}\\
\cos k_2e^{i\varphi}&0&\cos k_1e^{-i\varphi}\\
\cos k_3e^{-i\varphi}&\cos k_1e^{i\varphi}&0
\end{pmatrix},
\label{khama}
\end{align}
with $k_i=\vec{k}\cdot\vec{a}_i$, and $\vec{a}_1=(1,0)$, $\vec{a}_2=(1/2,\sqrt{3}/2)$, $\vec{a}_3=\vec{a}_2-\vec{a}_1$ are the lattice vectors.  Diagonalizing the Hamiltonian  gives three magnon branches of the kagom\'e ferromagnet. In the following we set $B=0$ as it simply shifts the magnon bands to high energy. As shown in Fig.~\eqref{klat}c, without the DM interaction, i.e. $D/J=0$ or $\varphi=0$, the two lower dispersive bands form Dirac magnon cones at $\pm{\bf K}$ (see Fig.~\eqref{klat}b), whereas the flat band has the highest energy and touches one of the dispersive bands quadratically at ${\bf \Gamma}$. In Fig.~\eqref{klat}d, we include a small DM interaction $D/J=0.15$ applicable to Cu(1-3, bdc) \cite{alex5b}.  Now the flat band acquires a dispersion and all the bands are separated by a finite energy gap with well-defined Chern numbers. Thus, the system becomes a topological magnon insulator \cite{zhh,alex4,alex5b,alex5c}. 

\vspace{10px}

\noindent\textbf{ Periodically driven topological magnon insulators.}~~ In this section, we introduce the notion of periodically driven intrinsic topological magnon insulators. Essentially, this concept will be based on the charge-neutrality of magnons in combination with their  magnetic dipole moment $\vec{ \mu} =g\mu_B\hat{z}$. Let us suppose that  magnons in insulating quantum magnetic systems are exposed to an  electromagnetic field with a dominant time-dependent electric field vector $\vec E(\tau)$. Then the effects of the field on the system can be described by a vector potential defined as $\vec E(\tau)=-\partial \vec A(\tau)/\partial \tau$, where 
\bea 
\vec A(\tau)=[-A_x\cos(\omega \tau+\phi), A_y\cos(\omega \tau),0],
\label{eqn4a}
\eea 
with amplitudes $A_x$ and $A_y$, frequency $\omega$, and phase difference $\phi$. The vector potential has time-periodicity: $\vec A(\tau+T)=\vec A(\tau)$, with $T=2\pi/\omega$ being the period.  Here $\phi=\pi/2$ corresponds to circularly-polarized light, whereas $\phi=0$ or $\pi$ corresponds to linearly-polarized light.

Using the AC effect for charge-neutral particles \cite{aha}, we consider magnon quasiparticles with magnetic dipole $\mu$ moving in the  background of a time-dependent electric field. In this scenario they will acquire a time-dependent AC phase (see Methods)  given by 
 \bea \theta_{\ell\ell^\prime}(\tau)=\mu_m\int_{\vec{r}_{\ell}}^{\vec{r}_{\ell^\prime}} \vec \Xi(\tau)\cdot d\vec{\ell},
 \label{eqn4}
 \eea 
 where $\mu_m = g\mu_B/\hbar c^2$ and $\vec{r}_\ell$ is the coordinate of the lattice at site $\ell$. We have used the notation $ \vec \Xi(\tau) = \vec{E}(\tau)\times \hat z$  for brevity. From Eq.~\eqref{eqn4a} we obtain 
 \begin{align}
 \vec \Xi(\tau)=[E_y\sin(\omega \tau),E_x\sin(\omega \tau+\phi),0],
 \end{align}
 where $E_x$ and $E_y$ are the amplitudes of the electric field and $\phi$ is the phase factor. By virtue of the time-dependent Peierls substitution,  the periodically driven magnon Hamiltonian is succinctly given by

\begin{align}
\mathcal H(\tau)&=-t_0\sum_{\la \ell\ell^\prime \ra} \big(e^{i[\varphi_{\ell\ell^\prime} +\theta_{\ell\ell^\prime}(\tau)]} a_{\ell}^\dagger a_{\ell^\prime}+ \text{H.c}\big) +t_z\sum_{\ell}n_\ell.
\label{kham1}
\end{align}
Therefore, the  time-dependent momentum space Hamiltonian $\mathcal H(\vec{k},\tau)$ corresponds to making the  time-dependent Peierls substitution  $\vec{k}\to \vec{k} +\mu_m\vec \Xi(\tau)$  in Eq.~\eqref{khama}. We note that previous studies based on the AC effect in insulating magnets considered  a time-independent electric field gradient, which leads to magnonic Landau levels \cite{mei1, mei, xr1,xr}. In stark contrast to those studies, the time-dependent version can lead to  Floquet topological magnon insulators in insulating quantum magnets with inversion symmetry, e.g. the honeycomb lattice \cite{ow1} or the Lieb lattice   (see Supplementary Information). We note that  Floquet topological magnon insulators can also be generated by driving a gapped trivial magnon insulator with vanishing Chern number, in a similar manner to Dirac magnons. A comprehensive study of this case is beyond the purview of this report.   In the current study, however, the kagom\'e lattice quantum  ferromagnets naturally lack inversion symmetry, and thus allows an intrinsic DM interaction as depicted in Fig.~\eqref{klat}a.
  \begin{figure}
\includegraphics[width=6in]{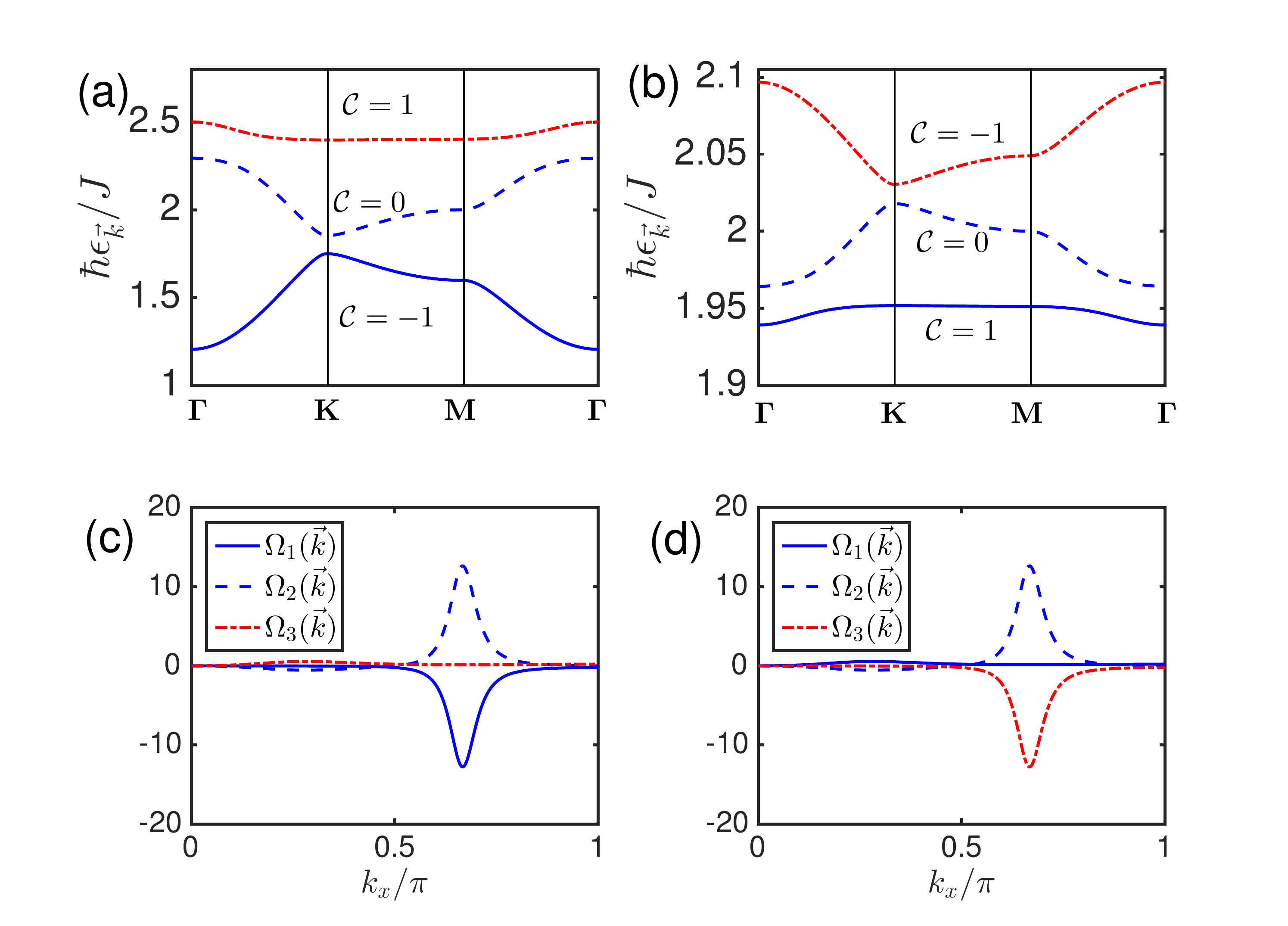}
\caption{ Top panel. Topological magnon bands of periodically driven topological magnon insulator at $D/J=0.15$. (a) $\mathcal{E}_x=\mathcal{E}_y=1.7$ and $\phi=\pi/2$. (b) $\mathcal{E}_x=\mathcal{E}_y=2.5$ and $\phi=\pi/2$. Bottom panel. Tunable Berry curvatures of periodically driven intrinsic topological magnon insulator on the kagom\'e lattice at $k_y=0$ and $D/J=0.15$. (c) $\mathcal{E}_x=\mathcal{E}_y=1.7$ and $\phi=\pi/2$. (d) $\mathcal{E}_x=\mathcal{E}_y=2.5$ and $\phi=\pi/2$.}
\label{BC}
\end{figure}

 \begin{figure}
\includegraphics[width=5.5in]{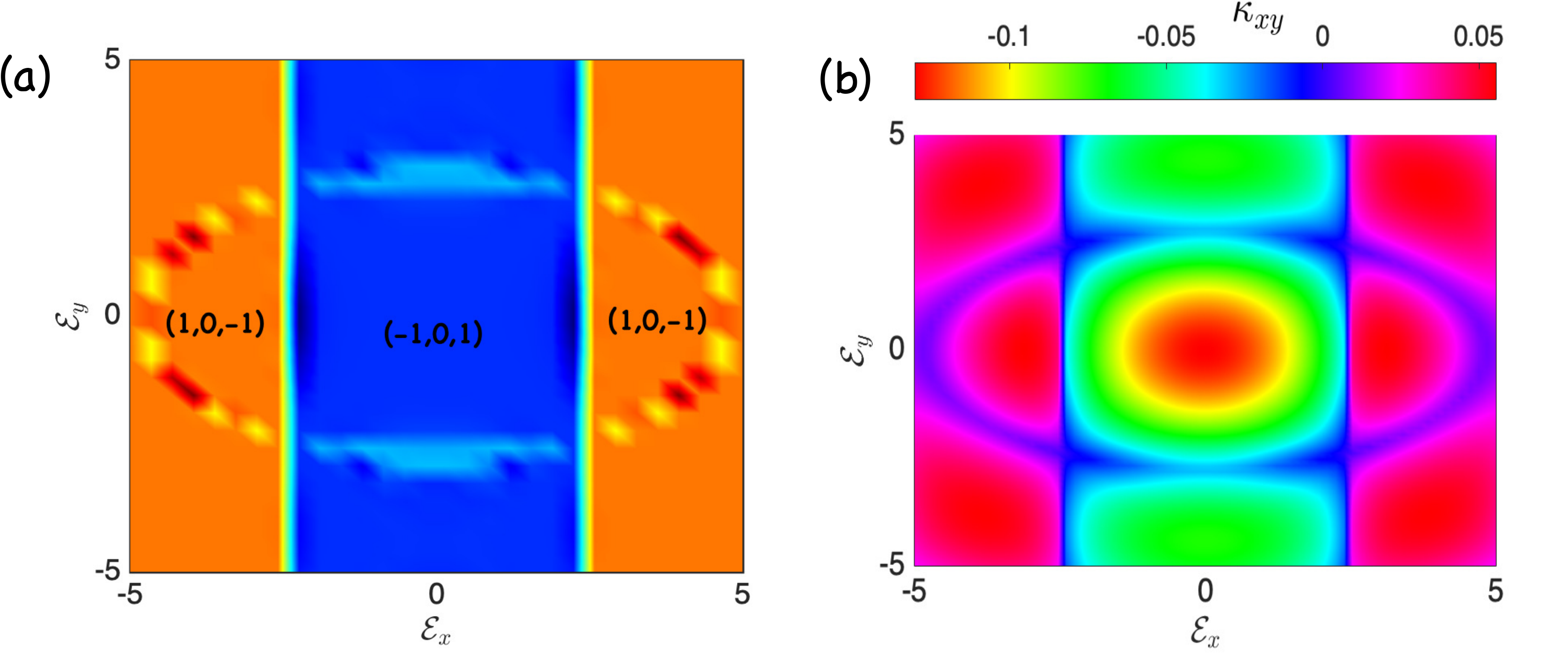}
\caption{Topological phase diagram of periodically driven intrinsic topological magnon insulator on the kagom\'e lattice. (a) Chern number phase diagram  for $D/J=0.15$ and $\phi=\pi/2$. (b) Thermal Hall conductivity phase diagram  for $D/J=0.15$, $\phi=\pi/2$, and $T=0.75$.}
\label{Phase}
\end{figure}

Now, we apply the  magnonic Floquet-Bloch theory in Methods. For simplicity, we consider the magnonic Floquet Hamiltonian in the off-resonant regime, when the driving frequency $\omega$ is larger than the magnon bandwidth $\Delta$ of the undriven system, i.e. $\omega\gg\Delta$. In this limit, the Floquet bands are decoupled, and it suffices to consider the zeroth order  time-independent Floquet magnon Hamiltonian  $\mathcal H^0(\vec{k})=t_z \text{I}_{3\times 3}- \Lambda^0(\vec{k}) $, where

\begin{align}
\Lambda^0(\vec{k}) &=
\begin{pmatrix}
0& t_0^{AB}\cos k_2e^{-i\varphi}& t_0^{CA}\cos k_3e^{i\varphi}\\
t_0^{AB}\cos k_2e^{i\varphi}&0&t_0^{BC}\cos k_1e^{-i\varphi}\\
t_0^{CA}\cos k_3e^{-i\varphi}&t_0^{BC}\cos k_1e^{i\varphi}&0
\end{pmatrix},
\label{kflo}
\end{align}
and
\begin{align}
t_0^{AB}&=2t_0\mathcal J_{0}\lb \frac{1}{2}\sqrt{\mathcal{E}_x^2+3\mathcal{E}_y^2+2\sqrt{3}\mathcal{E}_x\mathcal{E}_y\cos\phi}\rb,\\ 
t_0^{BC}&=2t_0\mathcal J_{0}\lb|\mathcal{E}_x|\rb,\\
t_0^{CA}&=2t_0\mathcal J_{0}\lb \frac{1}{2}\sqrt{\mathcal{E}_x^2+3\mathcal{E}_y^2-2\sqrt{3}\mathcal{E}_x\mathcal{E}_y\cos\phi}\rb,
\end{align}
where $\mathcal J_n$ is the Bessel function of order $n$.  The dimensionless quantity characterizing the intensity of light is different from that of electronic systems and it is given by
\begin{align}
\mathcal{E}_i =\frac{g\mu_B E_i a}{\hbar c^2},
\end{align}
where $ \quad  i = x,y$ and $a$ is the lattice constant.
Evidently, a direct consequence of photo-irradiation  is that the  magnonic Floquet Hamiltonian \eqref{kflo} is equivalent to that of a distorted kagom\'e ferromagnet with unequal tunable interactions $t_0^{AB}\neq t_0^{BC}\neq t_0^{CA}$. In the following, we shall discuss the topological aspects of this model. The Berry curvature is one of the main  important quantities in topological systems. It is the basis of many observables in topological insulators. To study the photoinduced topological phase transitions in driven topological magnon insulators, we define the Berry curvature of a given magnon band $\alpha$   as
\begin{align}
\Omega_{\alpha}(\vec k)=-2\text{Im}\sum_{\alpha^\prime \neq \alpha}\frac{\big(\braket{\psi_{\vec k, \alpha}|\hat v_x|\psi_{\vec k, \alpha^\prime}}\braket{\psi_{\vec k, \alpha^\prime}|\hat v_y|\psi_{\vec k, \alpha}}\big)}{\big(\epsilon_{\vec k, \alpha}- \epsilon_{\vec k, \alpha^\prime}\big)^2},
\label{chern2}
\end{align}
where $\hat v_{x,y}=\partial \mathcal{H}^0(\vec k)/\partial k_{x,y}$ are the velocity operators,  $\psi_{\vec k, \alpha}$ are the magnon eigenvectors, and  $\epsilon_{\vec k, \alpha}$ are the  magnon energy bands. The associated Chern number is defined as the integration of the Berry curvature over the Brillouin zone (BZ),
\begin{align}
\mathcal C_\alpha=\frac{1}{2\pi}\int_{BZ} d^2 k~\Omega_{\alpha}(\vec k),
\end{align}
where $\alpha=1,2,3$ label the lower, middle, and upper magnon bands respectively.

 In Fig.~\ref{BC} we have shown the evolution of the magnon bands and the Berry curvatures for varying  light intensity. We  can see that the lower and upper magnon bands and their  corresponding Berry curvatures   change  with varying light intensity, whereas the middle magnon band remains unchanged. Consequently, the system   changes from one topological magnon insulator with Chern numbers $(-1,0,1)$ to another one with Chern numbers $(1,0,-1)$ as shown in the photoinduced topological phase diagram in Fig.~\ref{Phase} (a).   In other words, exposing a topological magnon insulator to a varying light intensity field redistribute  the magnon band structures and subsequently leads to a topological phase transition between one topological magnon insulator to another with different Berry curvatures and Chern numbers. 

A crucial consequence of topological magnon insulators is the thermal Hall effect \cite{alex5a, alex1}.   Theoretically,  the thermal Hall effect is understood as a consequence of the Berry curvatures induced by the DM interaction \cite{alex2, alex0, alex2q}. If we focus on the regime in which the Bose distribution function is close to equilibrium, the same theoretical concept of undriven thermal Hall effect can be applied in the photoinduced system.  The transverse component $\kappa_{xy}$ of the thermal Hall conductivity is given explicitly as \cite{alex2, alex2q}
\begin{align}
\kappa_{xy}=-k_B^2 T\int_{BZ} \frac{d^2k}{(2\pi)^2}~ \sum_{\alpha=1}^N c_2\lb n_\alpha\rb\Omega_{\alpha}(\vec k),
\label{thm}
\end{align}
where $ n_\alpha=n\big[ \epsilon_{\alpha}(\vec k)\big]=1/ \big[e^{\epsilon_{\alpha}(\vec k)/k_BT}-1\big]$ is the Bose distribution function close to thermal equilibrium,  $k_B$ is the Boltzmann constant, $T$ is the temperature, and $ c_2(x)=(1+x)\lb \ln \frac{1+x}{x}\rb^2-(\ln x)^2-2\text{Li}_2(-x)$, with $\text{Li}_2(x)$ being the  dilogarithm. Indeed, the thermal Hall conductivity is the Berry curvature weighed by the $c_2$ function. Therefore, any change of the Berry curvature will affect the thermal Hall conductivity.  Evidently, as shown in Fig.~\ref{Phase} (b), the two photoinduced phases of the intrinsic topological magnon insulator have different signs of the anomalous thermal Hall conductivity due to the sign change in the Berry curvatures. The elliptic ring in the topological phase diagram in Fig.~\ref{Phase} is an artifact of the kagom\'e lattice, together with circularly polarized light. It does not exist  with linearly polarized light, and it is also not present  on the honeycomb lattice.

 \begin{figure*}
\includegraphics[width=7in]{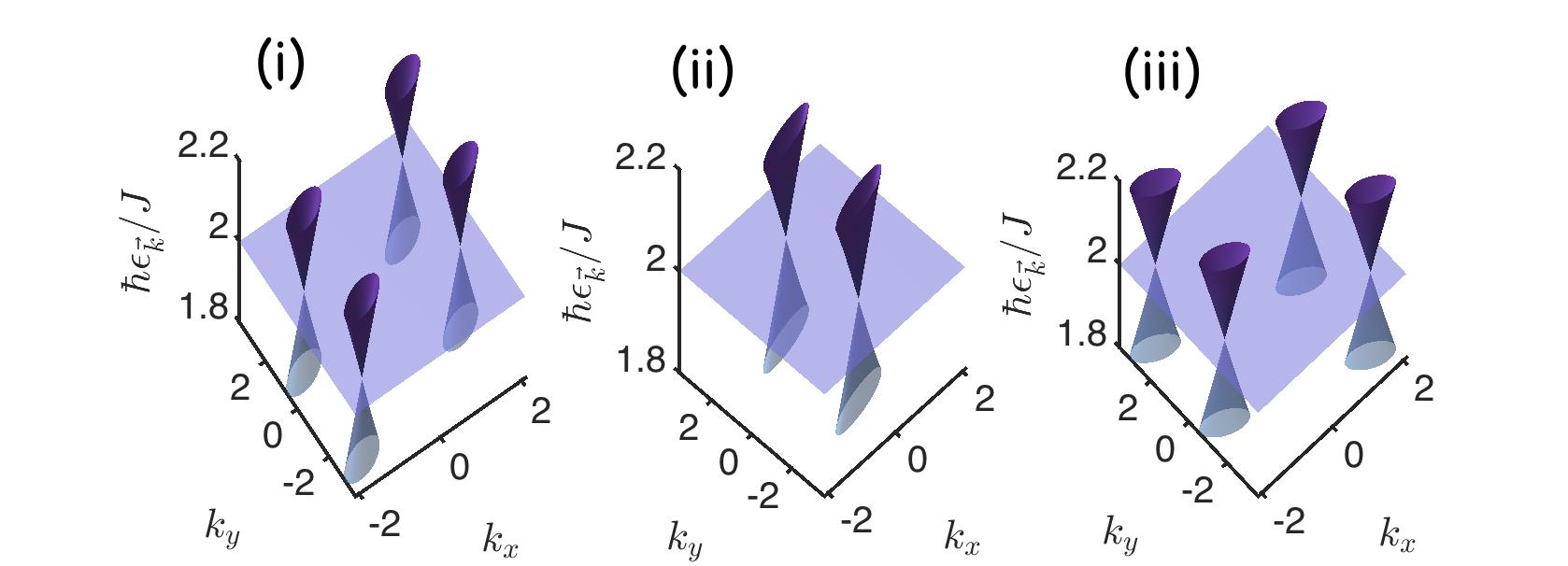}
\caption{Photoinduced pseudospin-1 topological magnon semimetals in periodically driven intrinsic topological magnon insulator on the kagom\'e lattice. (i) $t_0^{AB}=0;~t_0^{BC}\neq t_0^{CA}\neq 0$ with $\mathcal{E}_x=1.7$, $\mathcal{E}_y=1.5$ and $\phi=\pi/2$. (ii) $t_0^{BC}=0;~t_0^{AB}\neq t_0^{CA}\neq 0$ with $\mathcal{E}_x=1.7$, $\mathcal{E}_y=2.5$ and $\phi=0$. (iii) $t_0^{CA}=0;~t_0^{BC}\neq t_0^{CA}\neq 0$ with $\mathcal{E}_x=1.7$, $\mathcal{E}_y=1.5$ and $\phi=\pi/2$. Here we set  $D/J=0.15$.}
\label{F_band}
\end{figure*}  

\vspace{10px}
\noindent\textbf{ Photoinduced topological magnon semimetal.}~~ The topological phase transitions in periodically driven intrinsic topological magnon insulators can also be extended to synthetic topological magnon semimetals with gapless magnon bands. As we mentioned above the photoinduced distorted interactions $t_0^{AB}\neq t_0^{BC}\neq t_0^{CA}$ can be controlled by the amplitude and the polarization of the light intensity, therefore there is a possibility to obtain new interesting magnon phases  in periodically driven intrinsic topological magnon insulators. Let us consider three different limiting cases of the photoinduced distorted interactions. 

(i): $t_0^{AB}=0;~t_0^{BC}\neq t_0^{CA}\neq 0$, which leads to the magnon bands  $\epsilon_{\vec k}^0=t_z$ and
\begin{align}
\epsilon_{ \vec k}^{\pm}&=t_z  \pm \frac{1}{\sqrt{2}}\sqrt{\lb t_0^{BC}\rb^2\big[ 1+\cos(2k_1)\big]+\lb t_0^{CA}\rb^2\big[ 1+\cos(2k_3)\big]}
\end{align}
 
 (ii): $t_0^{BC}=0;~t_0^{AB}\neq t_0^{CA}\neq 0$. The magnon bands in this case are given by $\epsilon_{\vec k}^0=t_z$ and
\begin{align}
\epsilon_{ \vec k}^{\pm}&=t_z  \pm \frac{1}{\sqrt{2}}\sqrt{\lb t_0^{AB}\rb^2\big[ 1+\cos(2k_2)\big]+\lb t_0^{CA}\rb^2\big[ 1+\cos(2k_3)\big]}
\end{align}

 (iii): $t_0^{CA}=0;~t_0^{BC}\neq t_0^{CA}\neq 0$. In this case we have  $\epsilon_{\vec k}^0=t_z$ and
\begin{align}
\epsilon_{ \vec k}^{\pm}&=t_z  \pm \frac{1}{\sqrt{2}}\sqrt{\lb t_0^{AB}\rb^2\big[ 1+\cos(2k_2)\big]+\lb t_0^{BC}\rb^2\big[ 1+\cos(2k_1)\big]}
\end{align}
In each case there are three magnon bands featuring  one flat magnon band and two dispersive magnon bands, similar to the undriven topological magnon insulator in Fig.~\eqref{klat}d.  However, in the present case there is a possibility to obtain other interesting magnon phases different from the gapped topological magnon bands in the undriven system.  For instance, cases (i)--(iii)  realize  pseudospin-1 Dirac-Weyl magnon cones or three-component bosons at ${\bf K}_1=\lb \pm \pi/2,\mp \pi/2\sqrt{3}\rb$, ${\bf K}_2=\lb 0,\mp \pi/\sqrt{3}\rb$, and ${\bf K}_3=\lb \pm \pi/2,\pm \pi/2\sqrt{3}\rb$  respectively, as indicated by red and green dots in Fig.~\ref{klat}(b). The pseudospin-1 Dirac-Weyl magnon cones occur at the  energy of the flat band $\epsilon_{{\bf K}_i}=t_z$ as shown in Fig.~\eqref{F_band}.   Expanding the Floquet-Bloch magnon Hamiltonian in the vicinity of ${\bf K}_1$ yields 
\begin{align}
\mathcal H^{0}({\bf K}_1 +{\vec q}) \simeq t_z \text{I}_{3\times 3} \pm v_xq_x\lambda_x \mp v_yq_y\lambda_y,
\end{align}
where $v_x= t_0^{BC}$ and $v_y =t_0^{CA}$. The ${\lambda}'s$ are  the  pseudospin-$1$ representation of the SU(2) Lie algebra $[\lambda_i, \lambda_j]=i\epsilon_{ijk}\lambda_k$, where 
\begin{align}
\lambda_x &=
\begin{pmatrix}
0& 0& e^{i\varphi}\\
0&0&0\\
e^{-i\varphi}&0&0
\end{pmatrix},~
\lambda_y =
\begin{pmatrix}
0& 0& 0\\
0&0&e^{-i\varphi}\\
0&e^{i\varphi}&0
\end{pmatrix},~\lambda_z =
\begin{pmatrix}
0& ie^{2i\varphi}& 0\\
-ie^{-2i\varphi}&0&0\\
0&0&0
\end{pmatrix}.
\end{align}
Similar pseudospin-$1$ linear Hamiltonian can be obtained for the Dirac-Weyl magnon cones around ${\bf K}_2$ and ${\bf K}_3$.

\vspace{10px}
\noindent\textbf{Conclusion}

\noindent We have presented a study of photoinduced topological phase transitions in periodically driven intrinsic topological magnon insulators. The main result of this report is that intrinsic topological magnon insulators in the kagom\'e ferromagnets can be driven to different topological phases with different Berry curvatures  using photo-irradiation. Therefore, each topological phase is associated with a different sign of the thermal Hall conductivity, which results in a sign reversal of the magnon heat photocurrent. These topological transitions require no external magnetic field.  Interestingly, we observed that by varying the light intensity, the periodically driven intrinsic topological magnon insulators can also realize synthetic gapless topological magnon semimetals with pseudospin-$1$ Dirac-Weyl magnon cones. We believe that our results should also apply to 3D topological magnon insulators. In fact,  a 3D topological magnon insulator should also have a Dirac magnon cone on its 2D surface, which can be photo-irradiated to engineer a 2D topological magnon insulator in analogy to electronic systems \cite{dik}.  Here, we have studied the off-resonant regime, when the driving frequency $\omega$ is larger than the magnon bandwidth $\Delta$ of the undriven system. In this regime, the Floquet sidebands are decoupled and can be considered independently. By lowering the driving frequency below the magnon bandwidth, the Floquet sidebands overlap, which results in photon absorption. In this limit the system would have several overlapping   topological phases depending on the polarization of the light. In general, we believe  that the predicted results in this report are pertinent to experiments and will remarkably impact future research in topological magnon insulators and their potential practical applications to  photo-magnonics \cite{benj} and magnon spintronics \cite{magn, benja}.

\vspace{10px}
\noindent\textbf{\large Methods}

\noindent\textbf{Neutral particle with magnetic dipole moment in an external electromagnetic field.}~~ Two-dimensional topological magnon insulators (or Dirac magnons) can be captured by  massive (or massless) (2+1)-dimensional Dirac equation  near $\pm {\bf K}$.   In general, a massive neutral particle with mass ($m$) couples non-minimally to an external electromagnetic field  (denoted by the tensor $ F_{\mu\nu}$) via its magnetic dipole moment ($\mu$). In (3+1) dimensions, the system is described  by the Dirac-Pauli Lagrangian \cite{bjo} 
\begin{align}
\mathcal L=\bar\psi(x)(i\gamma^\mu\partial_\mu-\frac{\mu_m}{2}\sigma^{\mu\nu} F_{\mu\nu}-m)\psi(x),
\end{align}
where $\hbar=c=1$ has been used. Here  $x\equiv x^\mu=(x^0,\vec x)$, $\bar\psi(x)=\psi^\dg(x)\gamma^0$, and $\gamma^\mu=(\gamma^0,\vec\gamma)$ are the $4\times 4$  Dirac matrices that obey the algebra \bea \lbrace \gamma^\mu,\gamma^\nu\rbrace=2g^{\mu\nu},~\text{where}~\quad g^{\mu\nu}=\text{diag}(1,-1,-1,-1),\eea and  \bea\sigma^{\mu\nu}=\frac{i}{2}[\gamma^\mu,\gamma^\nu]=i\gamma^\mu\gamma^\nu,\quad (\mu\neq \nu).\eea 

For the purpose of our study in this report, we consider an electromagnetic field with only  spatially uniform and time-varying electric field vector $\vec{E}(\tau)$ (however, the resulting AC phase is valid for a general electric field  $\vec{E}(\tau,\vec r)$). In this case, the corresponding Hamiltonian is given by
\begin{align}
\mathcal H=\int d^3 x ~\psi^\dg(x)\big[\vec{\alpha}\cdot\big(-i\vec{\nabla}-i\mu_m\beta\vec{E}(\tau)\big)+m\beta\big]\psi(x),
\label{eqn21}
\end{align}
where $\vec{\alpha}=\gamma^0\vec \gamma$, and $\beta=\gamma^0$. 

In (2+1) dimensions,  the Hamiltonian \eqref{eqn21} corresponds to that of 2D topological magnon insulators (near $\pm {\bf K}$) with  magnetic dipole moment $\vec \mu=\mu_m\hat z$.   In this case,  the Dirac matrices  are simply Pauli matrices given by
\begin{align}
\beta=\gamma^0=\sigma_z,~\gamma^1=i\sigma_y,~\gamma^2=-i\sigma_x.
\end{align}
The corresponding momentum space Hamiltonian in (2+1) dimensions now takes the form
\begin{align}
\mathcal H=\int \frac{d^2k}{(2\pi)^2}~\psi^\dg(\bo, \tau)\mathcal H(\bo, \tau)\psi(\bo, \tau),
\end{align}
where
\begin{align}
\mathcal H(\bo, \tau)=\vec{\sigma}\cdot\big[\bo+\mu_m\big(\vec{E}(\tau)\times \hat z\big)\big]+m\sigma_z,~\text{with}~\vec{\sigma}=(\sigma_x,\sigma_y).
\label{eqn24}
\end{align}
We can clearly see the time-dependent AC phase from the Hamiltonian in Eq.~\ref{eqn24}.

\vspace{10px}
\noindent\textbf{Magnonic Floquet-Bloch theory.}~~ Periodically driven quantum systems are best described by the Floquet-Bloch theory. The magnonic  version describes the interaction of light with magnonic Bloch states in insulating magnets. In this section, we develop this theory for the time-dependent magnon Hamiltonian Eq.~\eqref{kham1} in momentum space. We consider the time-dependent Schr\"{o}dinger equation for the system
\begin{align}
i\hbar\frac{d\ket{\psi(\vec{k},\tau)}}{d\tau}=\mathcal{H}(\vec{k},\tau)\ket{\psi(\vec{k},\tau)},
\end{align}
where $\ket{\psi(\vec{k},\tau)}$ is the driven wave function. Due to the periodicity of the vector potential $\vec A(\tau)$, the driven Hamiltonian $\mathcal H(\vec{k},\tau)$ is also periodic and can be expanded in Fourier space as
\begin{align}
\mathcal{H}(\vec{k},\tau)= \mathcal{H}(\vec{k}, \tau+T)=\sum_{n=-\infty}^{\infty} e^{in\omega \tau}\mathcal{H}_n(\vec{k}),
\end{align}
where $\mathcal{H}_n(\vec{k})=\frac{1}{T}\int_{0}^T e^{-in\omega \tau}\mathcal{H}(\vec{k}, \tau) d\tau=\mathcal{H}_{-n}^\dg(\vec{k})$ is the Fourier component. The ansatz for solution to the  Schr\"{o}dinger equation can be written as 
\begin{align}
\ket{\psi_\alpha(\vec{k}, \tau)}&=e^{-i \epsilon_\alpha(\vec{k}) \tau}\ket{\xi_\alpha(\vec{k}, \tau)}=e^{-i \epsilon_\alpha(\vec{k}) \tau}\sum_{n=-\infty}^{\infty} e^{in\omega \tau}\ket{\xi_{_\alpha,n}(\vec{k})}
\end{align}
where  $\ket{\xi_\alpha(\vec{k}, \tau)}$ is the time-periodic Floquet-Bloch wave function of magnons and $\epsilon_\alpha(\vec{k})$ are the magnon quasi-energies. The corresponding Floquet-Bloch eigenvalue equation is given by  $\mathcal{H}_F(\vec{k},\tau)\ket{\xi_{\alpha}(\vec{k},\tau)}=\epsilon_\alpha(\vec{k})\ket{\xi_{\alpha}(\vec{k},\tau)}$, where $\mathcal{H}_F(\vec{k},\tau)=\mathcal{H}(\vec{k},\tau)-i\partial_\tau$ is the Floquet operator. This leads  to a time-independent Floquet eigenvalue equation
\begin{align}
\sum_m [\mathcal H^{n-m}(\vec k) + m\omega\delta_{n,m}]\xi_{\alpha}^m(\vec{k})= \epsilon_\alpha(\vec{k})\xi_{\alpha}^n(\vec{k}),
\end{align}
where $\mathcal H^{p}(\vec k)=\frac{1}{T}\int_0^T d\tau e^{-ip\omega \tau}\mathcal H^{p}(\vec k, \tau)$. The associated Bessel function integral is given by

\begin{align}
\frac{1}{T}\int_0^T d\tau e^{-ip\omega \tau}e^{iz\sin(\omega\tau)}e^{iz^\prime\sin(\omega\tau+\phi)}=e^{ip\arctan \lb \frac{z^{\prime}\sin(\phi)}{z+z^{\prime}\cos(\phi)} \rb}\mathcal{J}_p\lb \sqrt{z^2+z^{\prime 2}+2zz^\prime\cos(\phi)}\rb,
\end{align}
where  $\mathcal J_p$ is the Bessel function of order $p$. In Eq.~\eqref{kflo} we consider the zeroth order approximation corresponding to $p=0$.

\vspace{10px}
\noindent\textbf{\large Acknowledgements}

\noindent Research at Perimeter Institute is supported by the Government of Canada through Industry Canada and by the Province of Ontario through the Ministry of Research
and Innovation. 

\vspace{10px}
\noindent\textbf{\large Author Contributions}

\noindent S. A. Owerre conceived the idea, performed the calculations, discussed the results, and wrote the manuscript.

\vspace{10px}
\noindent\textbf{\large Additional Information}

\noindent\textbf{Supplementary information} accompanies this paper.

\vspace{10px}
\noindent\textbf{Competing Interests}. I declare that the author has no competing interests as defined by Nature Research, or other interests that might be perceived to influence the results and/or discussion reported in this paper.
\end{document}